\documentclass[twocolumn,showpacs,fleqn,nobibnotes]{revtex4}
\usepackage{amsmath}
\usepackage{graphicx}
\usepackage{float}
\usepackage{subfigure}
\usepackage{psfig}

\begin{document}

%\pacs{ PACS numbers:14.65.Dw, 14.40.Lb, 13.85.Ni, 13.75.Ew}
%\narrowtext

\title{On the energy dependence of the  $D^+/D^-$ production asymmetry}
\author{E.R. Cazaroto$^1$, V.P. Goncalves$^2$, F.S. Navarra$^1$ and M. Nielsen$^1$}
\affiliation{$^1$Instituto de F\'{\i}sica, Universidade de S\~{a}o Paulo,
C.P. 66318,  CEP 05315-970, S\~{a}o Paulo, SP, Brazil\\
$^2$ Instituto de F\'{\i}sica e Matem\'atica,  Universidade
Federal de Pelotas\\
C. P. 354, CEP 96010-900, Pelotas, RS, Brazil\\}

%%%%%%%%%%%%%%%%%%%%%%%%%%%%%%%%%%%%%%%%%%%%%%%%%%%%%%%%%%%%
\begin{abstract}
In this paper we discuss the origin of the asymmetry present in  $D$  meson production and its energy dependence.  
In particular, we have applied the meson cloud model  to calculate the asymmetries in $D^-/D^+$ meson production in
high energy  $p-p$ collisions and find a good agreement with  recent LHCb data. 
Although small, this non-vanishing  asymmetry may shed light on the role played by the charm meson cloud 
of the proton.

\end{abstract}
%%%%%%%%%%%%%%%%%%%%%%%%%%%%%%%%%%%%%%%%%%%%%%%%%%%%%%%%%%%%%%



%%%%%%%%%%%%%%%%%%%%%%%%%%%%%%%%%%%%%%%%%%%%%%%%%%%%%%

\maketitle
%%%%%%%%%%%%%%%%%%%%%%%%%%%%%%%%%%%%%%%%%%%%%%%%%%%%%%

\section{Introduction}
\label{sec:Introdu}

It is experimentally well known  \cite{DATA1,DATA2,DATA3,DATA4,WA89_99,selex} that there is 
a significant difference between the  $x_F$ (Feynman momentum) distributions of $D^+$ and $D^-$   
mesons produced in hadronic collisions with proton, $\Sigma^-$ and 
pion projectiles. It is usually quantified in terms of the asymmetry function:
\begin{equation}
A = \frac{N_{D^-} - N_{D^+}} {N_{D^-} + N_{D^+}}
\label{assim_0}
\end{equation}
where $N$ may represent the number of mesons of a specific type or its distribution in $x_F$, rapidity $y$ and $p_T$. 
The  recent data of the COMPASS collaboration \cite{compass} have confirmed the 
existence of charm  production asymmetries also in $\gamma p $ collisions. Moreover, 
the very recent data from the LHCb collaboration \cite{DLHCb} showed that there is asymmetry in the production of 
$D^+$ and $D^-$ mesons in proton-proton collisions at $7$ TeV.
The origin of these asymmetries is still an open question.
It is not possible to understand these production asymmetries only with usual perturbative QCD (pQCD)  or with 
the string fragmentation model contained in PYTHIA.  This has motivated the construction of alternative models \cite{nnnt, icm1,icm2} 
which were able to obtain a reasonable description of the low energy data and make concrete  predictions for higher energy collisions.
The LHCb data allow us, for the  first time,  to compare the predictions of the models with 
high energy data. Moreover, studying the energy dependence of production asymmetries, it may be possible to learn more 
about forward charm production, which undoubtedly has a non-perturbative component \cite{icm1}. 
In this work we update one of these models, the meson cloud model or MCM \cite{nnnt}, and compare its predictions with the new LHCb data.

\begin{figure}[t]
\includegraphics[scale=0.25]{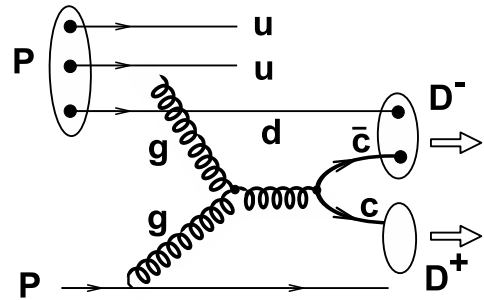}
\caption{ $D$ meson production in a proton - proton reaction.  The charm pair is created by gluon fusion and then the quarks
may fragment independently, as it happens to the $c$ quark in figure, or they can ``recombine" or ``coalesce" with one of the
projectile valence quarks, as it is the case of the $\bar{c}$ in the figure.  Mesons formed by recombination are harder since 
they are ``dragged" by the hard valence quark. }
\label{Fig. 1}
\end{figure}

Let us now  briefly review some ideas about charm production.
In perturbative QCD the most relevant elementary processes which are responsible for charm production are 
$q + \bar{q} \rightarrow c + \bar{c}$ and $g + g \rightarrow c + \bar{c}$. At high energies, due to the growth 
of the gluon distributions, the latter should be dominant.  In standard pQCD, 
after being produced  the $c$ and $\bar{c}$ quarks fragment independently and hence, the resulting mesons $D^+$ and $D^-$ 
(also $D^0$ and $ \bar{D}^0$) will have the same rapidity, $p_T$ and $x_F$ distributions. This is indeed true for the 
bulk of charm production. Differences between the $D^+$ and $D^-$  $x_F$ distributions appear at large $x_F$, with 
$D^-$ being harder. Given the valence quark content of the proton $p (u u d)$ and of the  $D^- ( d \bar{c})$, 
a natural explanation of the observed effect is that the $\bar{c}$  is dragged  by the projectile valence $d$ quark, 
forming the, somewhat harder, $ d \bar{c} $ bound state. This process has been usually called recombination or coalescence and it is
illustrated in Fig. \ref{Fig. 1}. This  is a non-perturbative process and recombination models have been first proposed long 
time ago \cite{chiu,hwa,russo}  and then used more recently \cite{newrec,rapp} to study the accumulated experimental data and 
to make predictions for the RHIC collisions. Unfortunately, the current  RHIC experimental set up did not allow for a precise 
determination of production asymmetries. However, a more detailed analysis of the heavy quark sector is expected to be possible in the upgraded RHIC facility - RHIC II \cite{rhic2}.

An alternative way to implement the idea of recombination is to use a purely hadronic 
picture of charm production, in which, instead of producing charm pairs and then recombining them with the valence quarks,
we assume that the incoming proton fluctuates into a virtual charm meson - charm baryon pair, which may be liberated during the 
interaction with the target. This kind of fluctuation is unavoidable in any field theoretical description of hadrons and, in fact, 
it was shown \cite{ku} to be quite relevant to the understanding of hadron structure. It has been also successfully applied to 
particle production in high energy soft hadron collisions \cite{hss,cdnn} and in \cite{nnnt} it has been extended to the charm sector. 
This mechanism, in which the ``meson cloud" plays a major role, is quite economical and can be 
improved systematically (see \cite{mel} for light mesons). From now on we shall call it meson cloud model (MCM). 
A simple and accurate description of charm asymmetry production at lower
energies ($\sqrt{s} \simeq 10 -40$ GeV) within the framework of the  MCM can be found in \cite{cdnn01}. 

Another popular model of forward charm production is the 
intrinsic charm model (ICM) \cite{icm1,icm2,gnu}. The existence of an intrinsic charm component in the wave function enhances 
charm meson production at large $x_F$. While IC, as formulated in \cite{icm1,icm2} is supported by several  phenomenological analises, it is, 
alone, not enough to explain the difference between the $D^+$ and $D^-$ $x_F$ distributions.  It is necessary to add a recombination mechanism 
(or ``coalescence") in this kind of model to account for the observed asymmetries.  In \cite{icm1,icm2} the intrinsic charm component of the proton 
is the higher Fock state $|u u d c \bar{c} \rangle$ and its existence is attributed to a multigluon fusion, which is not calculable in pQCD. In 
\cite{nnnt} the intrinsic charm component of the proton was a consequence of the meson-baryon fluctuations mentioned above. In this approach,
since they ``feel'' the virtual bound states where they once were, the $c$ and $\bar{c}$ distributions are different from the beginning and when 
they later undergo independent fragmentation their difference will be transmitted to the final $D$ mesons. Thus, in its ``meson cloud'' version, 
intrinsic charm may account for large $x_F$ charm meson production including the asymmetries. The charm quarks of the 
projectile wave function traverse 
the target and fragment independently. In this corner of the phase space this mechanism may be more effective than gluon fusion because the later 
generates final mesons with a $x_F$ distribution peaked at zero. This possibility was explored in \cite{gnu}. 

The appearance of the first LHCb data on asymmetries \cite{DLHCb}  opens the exciting possibility of studying the energy dependence  
of forward charm production. Indeed,  more than ten years from the last data on this subject, we have now data taken at an energy which is larger 
than the previous one by a factor of $\simeq 200$ !  What do the models discussed above have to say about the energy dependence of forward 
charm production? What will be the fate of the production asymmetries ?  The answer to this question is interesting not only to the hadron physics 
community but also to the studies of CP violation, since a change in the relative  yield of particles and antiparticles may affect the interpretation 
of their decays and hence  change the ammount of CP violation. 

Naively, we expect that perturbative processes grow faster and become more important than non-perturbative ones as the reaction energy increases. As 
a consequence the asymmetries would gradually disappear. In \cite{nosco} a kinematical treatment of this problem arrived at the conclusion that, as 
the collision energy grows, the energy deposition in the central region increases. Baryon stopping also increases, the remnant valence quarks emerge 
from the collision with  less energy and when they recombine with charm antiquarks the outgoing charm  mesons with these quarks will be decelerated   
and their $x_F$ distribution will become invisible, buried under the much higher contribution from
the (symmetric)  central gluon fusion. In \cite{rapp}, using a recombination approach, the authors concluded that the asymmetry 
remains nearly constant with energy. In the meson cloud approach of  \cite{cdnn01} the  energy dependence of the asymmetry $A$ is 
approximately given by: 
\begin{equation}
A \propto \frac{\sigma_{hp}(s)}{\sigma_{c \bar{c}}(s)}
\label{a_frac}
\end{equation}
where $\sigma_{hp}$ is hadron-proton cross section and $\sigma_{c \bar{c}}$ is the total $c \bar{c}$ pair
production cross section, which grows faster than the hadronic cross sections. Therefore the asymmetry decreases with the energy.

In this work we use the model developed in \cite{cdnn01}  to study the recent LHCb data on $D^+ / D^-$ asymmetry and 
to check if the energy behavior of this asymmetry can be satisfactorily understood with this  model. The paper is organized as follows. In the next Section, we discuss the asymmetry production in terms of the meson cloud model, presenting the main formulas and assumptions of  the model.  In Section \ref{res} we present our results for the asymmetries in $\Lambda$ and $D$ production at SELEX ($\sqrt{s} = 33$ GeV) and LHCb ($\sqrt{s} = 7$ TeV) energies. We compare the results with the current data and make predictions for $\sqrt{s} = 14$ TeV.  Finally, in Section \ref{sum} we summaryze our main conclusions.

\section{Asymmetry Production in the Meson Cloud Model}
\label{sec:asymmmcm}

\subsection{The interaction between the cloud and the target}

In the  MCM we assume that quantum fluctuations in the projectile play an  
important role. The proton    may be decomposed in a series of Fock states,
containing states such as $ |p\rangle = | u u d \bar{c} c \rangle$. In the 
MCM we write the Fock decomposition in terms of the equivalent hadronic states, such as 
$|p\rangle = |\Sigma_c^{++} D^-\rangle $. This expansion contains  the ``bare'' terms 
(without cloud  fluctuations),  light  states and 
states containing the produced charmed meson ($D$ or $D_s$). The latter are, 
of course very much suppressed but they will be responsible for asymmetries. 
The ``bare'' states occur with a higher probability and are responsible for 
the bulk of charm meson production at low and medium momentum ($x_F \leq 0.4$), 
including, for example the perturbative QCD contribution. The cloud states are 
less frequent fluctuations and contribute to $D$ production in the ways described 
below. More precisely we shall assume that: 
\begin{equation}
| p\rangle = Z \,[ \, |p_0\rangle \, + \, ... \, + \,| M B \rangle \, + \, ... \, + \, 
|\Lambda_c  \bar{D}_0\rangle \, + \, |\Sigma_c^{++} D^-\rangle ]
\label{decp}
\end{equation}
where $Z$ is a normalization constant, $|p_0\rangle$ is the ``bare'' proton and the 
``dots'' denote all possible meson (M) - baryon (B) cloud states 
$|M B \rangle$ in the proton. The 
relative normalization of these states is fixed once the cloud 
parameters are fixed. The proton   is thus regarded as being a sum 
of virtual meson-baryon  pairs and a  proton-proton reaction can thus 
be viewed as a reaction between the ``constituent'' mesons and baryons  of 
the  projectile proton  with the target proton (or nucleus).

\begin{figure}[t]
\includegraphics[scale=0.23]{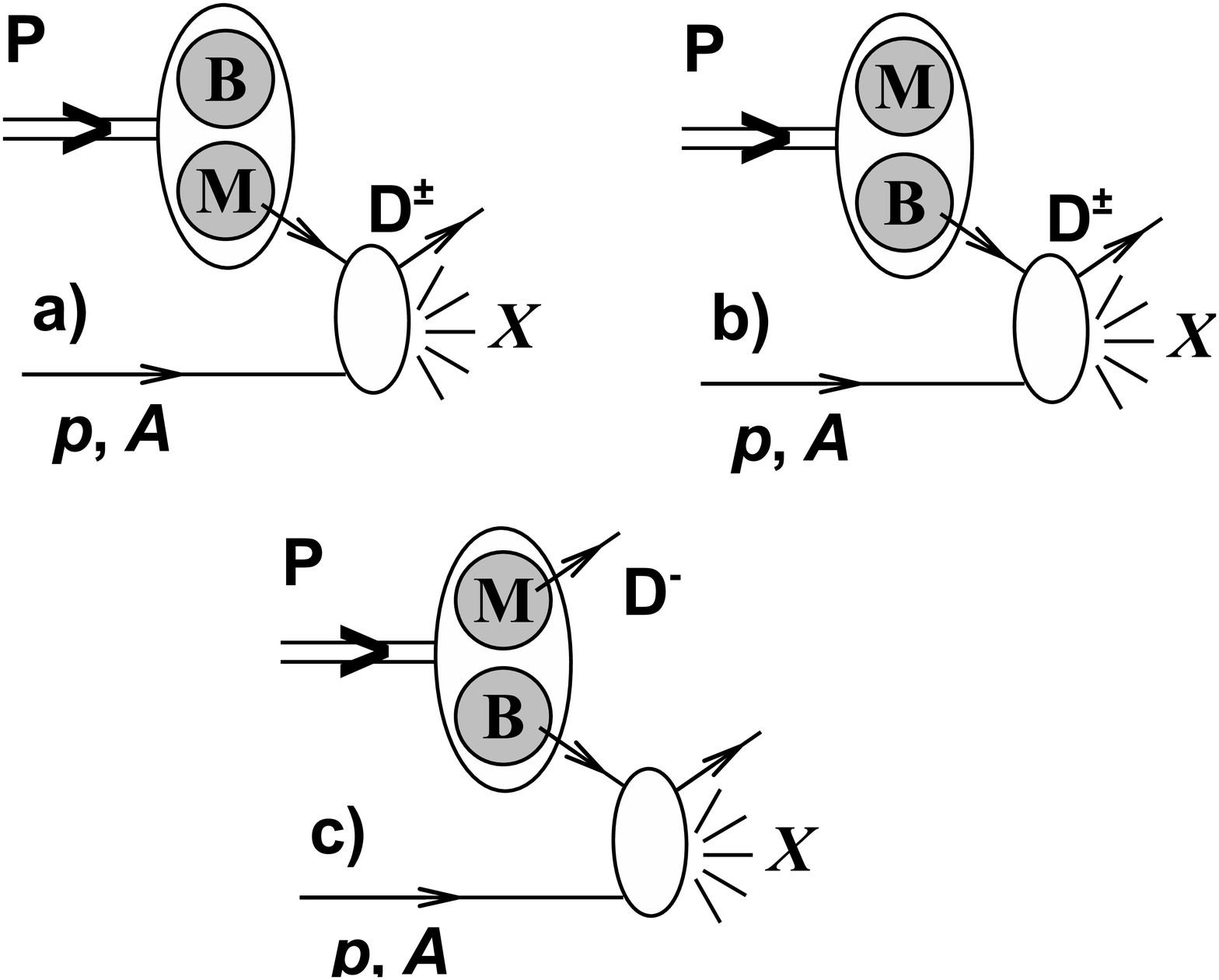}
%\vspace{01cm}
\caption{$p p$ collision in which the projectile is in a $|M B\rangle$ state. 
Figs. a) and b) show the ``indirect'' $D^\pm$  production and c) the
``direct'' $D^-$  production.}
\label{Fig. 2}
\end{figure}

With a proton beam the possible reaction mechanisms for $D^-$ meson production at large $x_F$ and small $p_T$ (the 
soft regime) are illustrated in Fig. \ref{Fig. 2}. 
In Fig. \ref{Fig. 2} a)  the baryon  just ``flies 
through'', whereas the corresponding meson interacts inelastically producing 
a $D$ meson in the final state. In Fig. \ref{Fig. 2} b)  the meson  just ``flies through'', 
whereas the corresponding baryon interacts inelastically producing a $D$ meson  
in the final state. In Fig. \ref{Fig. 2} c)  the meson in the cloud  {\it is already a $D^-$ } 
which escapes (similar considerations hold for $D^-$  production 
with a $\pi^-$ beam). This last mechanism is the main responsible for generating  
asymmetries. We shall refer to the  first two processes as  ``indirect 
production'' ($I$) and to the last one as ``direct production'' ($D$). The first 
two are calculated with  convolution formulas whereas the last one is given basically 
by the meson momentum distribution in the  initial $ |MB \rangle $ cloud state. 
Direct production has been widely used in the context of the MCM and applied to 
study $n$, $\Delta^{++}$ and $\pi^0$ production \cite{hss}. Indirect meson 
production has been considered previously in  \cite{cdnn}.

Inside the baryon, in the $ | MB \rangle $ state,  the meson and baryon 
have  fractional momentum $y_M$ and $y_B$ with distributions called  
$f_{M/MB}(y_M)$ and $f_{B/MB}(y_B)$ respectively (we shall use for them 
the short notation $f_M$ and $f_B$). Of course, by momentum conservation,  
$y_M + y_B = 1$ and these distributions are related by \cite{ku,cdnn}:
\begin{equation}
f_{M}(y) =  f_{B}(1-y) 
\label{fcomp}
\end{equation}
The ``splitting function'' $f_{M} (y)$ represents the probability density to find a meson 
with momentum fraction $y$ of the total cloud state $|MB\rangle$. With $f_M$ and 
$f_B$ we can compute the  differential  cross section for production of   $D¯$, $\bar{D}_0$ and 
$\Lambda_c$.  In what follows we write the formulas for the specific case of   $D¯$ production but 
it is easy, with the proper replacements, to write the corresponding expressions for $\bar{D}_0$ and
$\Lambda_c$. In  the reaction $p p \rightarrow  D^- X$   the  differential  cross section for   $D¯$ production  is given by:
\begin{equation}
\frac{ d \sigma^{p p \rightarrow D X}}{d x_F}   =\, \Phi_0 \,\, + \,\,  
\Phi_I\,\, +\,\, \Phi_D
\label{sechoque} 
\end{equation}
where $\Phi_0$ and $\Phi_I$  refer respectively to ``bare'' and indirect 
contributions to $D$ meson production and $x_F$ is  the fractional longitudinal 
momentum of the outgoing meson.  $\Phi_D$ represents the direct process depicted 
in Fig.  \ref{Fig. 2} c) and is given by \cite{hss,cdnn}:
\begin{equation}
\Phi_D =  \frac{\pi}{x_F} \, f_{D} (x_F) \, \sigma^{\Sigma p}
\label{direct}
\end{equation}
where $ f_{D} \equiv f_{D^- / \Sigma_c^{++} D^-}$ and $\sigma^{\Sigma p}$ is the total 
$p \, \Sigma_c^{++}$ cross section. 

In the MCM  the proton is from the start replaced by meson ($M$) and baryon ($B$) constituents, which interact independently with the 
target. In the projectile frame the $M$ and $B$ constituents can be considered as approximately free, since their interaction energy 
is much smaller than the energy carried by the incoming proton, which will smash $M$ or $B$ individually.   
This is sometimes called the impulse approximation. The subprocesses $M + \mbox{target} \rightarrow D^{\pm} + X$ and 
$B + \mbox{target} \rightarrow D^{\pm} + X'$ involve different initial and final states and their amplitudes are not supposed to be added 
(and subsequently squared).  We have rather to compute the corresponding 
cross sections, which we call $\sigma^{Mp}$ and $\sigma^{Bp}$, multiply them by the respective weigth, given by the function $f(y)$, and then 
sum the cross sections. This is why, in our case, the cross section reduces to the sum shown in Eq. (\ref{sechoque}).
The splitting function $f(y)$ comes already from a squared amplitude, it is positive definite and it is 
interpreted as a  probability \cite{ku}.

Replacing $D^{\pm}$ by $\Lambda_c / \bar{\Lambda}_c$ in
Figs. 2a and 2b, exchanging B with M and replacing $D^-$ by $\Lambda_c$ in Fig.  \ref{Fig. 2} c) we have a 
pictorial representation of $\Lambda_c$ production in the MCM with the following expression for 
the direct process:
\begin{equation}
\Phi_D =  \frac{\pi}{x_F} \, f_{\Lambda} (x_F) \, \sigma^{D p} 
\label{directla}
\end{equation}
where $ f_{\Lambda} \equiv f_{\Lambda_c / \Lambda_c \bar{D}_0}$ and $\sigma^{D p}$ is the total 
$\bar{D}_0  \, p$ cross section.
Analogous expressions can be written for the reaction $\pi^- p \rightarrow D X$. 

\subsection{The asymmetry}

Using (\ref{sechoque}), we can compute the cross sections and also the 
leading ($D^-$)/nonleading($D^+$) asymmetry:
\begin{eqnarray}
A^D(x_F) &=& \frac{ \frac{d \sigma^{D^-}(x_F)}{d x_F} \,\,-\,\, \frac{ d \sigma^{D^+}(x_F)}{d x_F}} 
{\frac{d \sigma^{D^-}(x_F)}{d x_F} \,\, + \,\, \frac{d \sigma^{D^+}(x_F)}{d x_F}}\nonumber \\
&=& \frac{\Phi_D \,\,+\,\,\Phi_I^{D^-}\,\,+\,\,\Phi_0^{D^-}\,\,-\,\,
\Phi_I^{D^+}\,\,-\,\,\Phi_0^{D^+}} {\Phi_D \,\,+\,\,\Phi_I^{D^-}\,\,+
\,\,\Phi_0^{D^-}\,\,+\,\,\Phi_I^{D^+} 
\,\,+\,\,\Phi_0^{D^+}}\nonumber \\
&\simeq& \frac{\Phi_D} {\Phi_D \,\,+\,\, 2 \, \Phi_I^{D}
\,\,+\,\, 2 \, \Phi_0^{D}} \equiv 
\frac{\Phi_D} {\Phi_T^D} 
\label{assym}
\end{eqnarray}
where the last line follows from assuming $\Phi_I^{D^-}=\Phi_I^{D^+}=\Phi_I^{D}$.   
This last assumption is made just 
for the sake of simplicity. In reality  
these contributions are not equal and their difference is an 
additional source of asymmetry, which we assume to be less important than $\Phi_D$. 
Since the ``bare'' states do not give origin to 
$D^- / D^+$ asymmetries (they represent mostly  perturbative QCD contributions
which rarely leave quark pairs in the large $x_F$ region), 
we have made use of  $\Phi_0^{D+} = \Phi_0^{D-} = \Phi_0^{D}$.  
The denominator of the above expression can be replaced by a parametrization of the  experimental data:
\begin{eqnarray}
\Phi_T^D &=& \frac{d \sigma^{D^-}(x_F)}{d x_F} \,\, + \,\, \frac{d \sigma^{D^+}(x_F)}{d x_F} \nonumber \\
         &=&  \sigma_0^D \, [ \, (1-x_F)^{n^-} \, + \,  (1-x_F)^{n^+} \, ] \nonumber \\
         &\simeq&  2 \, \sigma_0^D \, (1-x_F)^{n_{D}}
\label{sigdata}
\end{eqnarray}
where $n^-$ and $n^+$ are powers used by the different collaborations to fit 
their data. Typically $n^+=5$ and $n^-=3.5$, as suggested by the data analysis performed in \cite{DATA1,DATA2,DATA3,DATA4,WA89_99,selex}. 
For the sake of simplicity we shall assume that  $n^+ = n^- = n_{D} = 5$ for $D$ mesons. Integrating the above expression we obtain the total 
cross section for charged charm meson production $\sigma^{D^{\pm}} = {1}/{3} \,  \sigma_0^D $. Assuming isospin symmetry the charged and neutral  
($\sigma^{D^0}$) production cross sections are equal. Neglecting the contribution of other (heavier) charm states we can relate the $D$ meson production 
cross section to the total $c - \bar{c}$ production cross section, $\sigma_{c \bar{c}}$, in the following way: 
\begin{equation} 
\sigma^{D^{\pm}} = \sigma^{D^0} = \sigma^{D}  = \frac{1}{3} \sigma_0^D = \frac{1}{2} \sigma_{c \bar{c}} \,\,\,.
\label{secd}
\end{equation}
From the above relation we can extract the parameter $\sigma_0^D$ from the experimentally measured $\sigma_{c \bar{c}}$:
\begin{equation} 
\sigma^D_0 = \,  1.5 \,  \sigma_{c \bar{c}} \,\,\,.
\label{estsig}
\end{equation}
Inserting  (\ref{direct}) and (\ref{sigdata}) into (\ref{assym})
the asymmetry becomes:
\begin{equation}
A^D(x_F) =\frac{\pi \, \sigma^{\Sigma p} }{2 \,  \sigma_0^D} \,\, 
\frac{ f_{D}(x_F) } {x_F \,  (1-x_F)^{n_D} } \,\,\,.
\label{finass}
\end{equation}
An analogous development can be performed for the   $\Lambda_c/\bar{\Lambda}_c$ production  asymmetry, $A^{\Lambda} (x_F)$.  
In the analysis made by the SELEX collaboration, the differential cross section for $\Lambda_c$ production was parametrized as:
\begin{eqnarray}
\Phi_T^{\Lambda} &=& \frac{d \sigma^{\Lambda_c}(x_F)}{d x_F} \,\, + \,\, \frac{d \sigma^{\bar{\Lambda}_c}(x_F)}{d x_F} \nonumber \\
         &\simeq&  2 \, \sigma_0^{\Lambda} \, (1-x_F)^{n_{\Lambda}}
\label{sigdatalamb}
\end{eqnarray}
with $n_{\Lambda} \simeq 2$. Integrating the above expression we obtain the total cross section for $\Lambda_c$ and $\bar{\Lambda}_c$ 
production  $\sigma^{\Lambda} = {2}/{3} \,  \sigma_0^{\Lambda} $, which, according to \cite{gavai},  can be related to the 
total $c - \bar{c}$ production cross section, $\sigma_{c \bar{c}}$, in the following way:
\begin{equation} 
\sigma^{\Lambda} = \frac{2}{3}  \sigma_0^{\Lambda}  \simeq  0.1 \,   \sigma_{c \bar{c}}
\label{estsiglam}
\end{equation}
from where we finally have:
\begin{equation} 
\sigma_0^{\Lambda} =  \,  0.15 \,   \sigma_{c \bar{c}}  \,\,\,.
\label{estlamzer}
\end{equation}
Using (\ref{fcomp})  to obtain $f_{\Lambda}$ and then inserting (\ref{directla}) and (\ref{sigdatalamb}) into (\ref{assym})
the asymmetry $A^{\Lambda} (x_F)$ can be written as:
\begin{equation}
A^{\Lambda}(x_F) =\frac{\pi \, \sigma^{D p}}{2 \,  \sigma_0^{\Lambda}} \,\, 
\frac{ f_{\Lambda}(x_F) } {x_F \,  (1-x_F)^{n_{\Lambda}}}
\label{finassla}
\end{equation}
The behavior of (\ref{finass})  [ Eq. (\ref{finassla})] is controlled by the splitting function $f_D (x_F)$ [$f_{\Lambda}(x_F)$]. 
\subsection{The  splitting function}
\label{sullivan}
We now  write the splitting function in the Sullivan  approach \cite{ku,cdnn}.
The fractional momentum distribution of a pseudoscalar meson $M$  in the state 
$|M B'\rangle$ (of a baryon  $|B \rangle$) is given by \cite{ku,cdnn}:
\begin{eqnarray}
f_{M} (y) &=& \frac{g^2_{MBB'}}{16 \pi^2} \, y 
\int_{-\infty}^{t_{max}}
dt \, \frac{[-t+(m_{B'}-m_{B})^2]}{[t-m_{M}^2]^2} \nonumber \\
&\times& F_{MBB'}^2 (t)
\label{ftho}
\end{eqnarray}
where $t$ and $m_{M}$ are the four momentum square and the mass of
the meson in the cloud state and $t_{max} = m^2_B \, y- m^2_{B'} \,  y/(1-y)$ 
is the maximum $t$, with $m_B$ and $m_{B'}$ respectively the $B$ and $B'$ masses. 
Following a phenomenological approach, we use for the baryon-meson-baryon form factor 
$F_{MBB'}$, the exponential form:
\begin{equation}
F_{M B B'} (t) =  \exp  \left( \frac{ t - m_{M}^2}{\Lambda^2_{M B B'}} \right)
\label{eq:form}
\end{equation}
where $\Lambda_{M B B'}$ is the  form factor cut-off parameter. 
%In the above equations $t$ and $m_{M}$ are the four momentum square and the mass of
%the meson in the cloud state and $t_{max}$ is the maximum $t$ given by:
%\begin{equation}
%t_{max} = M^2_B \, y- \frac{M^2_{B'} \,  y}{1-y}  \,\,\,\, , 
%\label{tmax1}
%\end{equation}
%where $M_B$ and $M_{B'}$ are respectively the $B$ and $B'$ masses. 
Considering the
particular case where $B = p$,  $B' = \Sigma_c^{++}$ and $M = D^-$, we   
insert (\ref{ftho}) into
(\ref{finass}) to obtain the final expression for the asymmetry in our approach:
\begin{eqnarray}
A^D(x_F) &=& \frac{N^D}{( 1 - x_F )^{n_D}}
\nonumber \\ 
&\times& \int_{-\infty}^{t_{max}}
dt \, \frac{[-t+(m_{\Sigma} - m_{p})^2]}{[t-m_{D}^2]^2}\,
F_{ p D \Sigma }^2 
\label{finassytho}
\end{eqnarray}
where 
\begin{equation}
 N^D = \frac{g^2_{p D \Sigma_c}  \, \sigma^{\Sigma_c p}}{ 32 \, \pi  \, \sigma_0^D}
\label{defn2}
\end{equation} 
With the replacements  $ g_{p D \Sigma_c} \rightarrow    g_{p D \Lambda_c}$ and $\sigma^{\Sigma_c p} \rightarrow \sigma^{\Lambda_c p} $ 
this same expression holds for the process $p p \rightarrow \bar{D}_0 X$. 
With the replacements  $ g_{p D \Sigma_c} \rightarrow    g_{p D \Lambda_c}$,  $\sigma^{\Sigma_c p} \rightarrow \sigma^{D p} $ 
and  $ \sigma_0^D \rightarrow  \sigma_0^{\Lambda}$ the above expression holds for the process $p p \rightarrow {\Lambda}_c X$. 
For the pion beam,  we need also the  splitting function of the state 
$|\pi^-> \, \rightarrow \, |D^{0 *} D^-\rangle$. In this state,
the $D$ meson  momentum distribution 
turns out to be identical to (\ref{ftho}) except for the
bracket in the numerator which takes the form   
$[-t + \left( (m_{\pi}^2 - m_{D^{0*}}^2 -t)/2 m_{D^{0*}} \right)^2 ]$, and for 
trivial changes
in the definitions, i.e., $g^2_{MBB'} \, \rightarrow g^2_{\pi D D^{0*}}$, 
$F_{MBB'} \, \rightarrow F_{\pi D D^{0*}}$ and 
$\Lambda_{M B B'} \, \rightarrow \, \Lambda_{\pi D D^{0*}}$.  Realizing that $y = x_F$ 
in the above equations, we can see that in the limit $x_F \rightarrow 1$, $\,$
$t_{max} \rightarrow - \infty$ and the integral in (\ref{finassytho}) goes to zero. In 
fact, it vanishes faster than the denominator and therefore $A \rightarrow 0$. This  
behavior does not depend on the cut-off parameter but it depends on the choice of the 
form factor. For a monopole form factor we may obtain asymmetries which grow even at 
very large $x_F$. 
%Since $t$ controls the off-shellness of the emitted meson, which, 
%in turn, is related to the virtuality of the $|M B' \rangle $ state (or $|M M' \rangle$ state in the 
%case of the pion beam), the vanishing of $A_{2}$ happens because aqui

To conclude this section we would like to point out that our calculation is based on quite general and well 
established ideas, namely that hadron projectiles fluctuate into hadron-hadron
(cloud) states and that these states interact with the target.  During the derivation of the expressions  for the 
asymmetry many strong assumption have been made. In the end our results 
depend  on two  parameters:  $\Lambda$ and $N$.  Whereas $\Lambda$ affects  the width
and position of the maximum of the momentum distribution of the leading  meson in
the cloud (and consequently of the asymmetry),  $N$ is a  multiplicative
factor which determines the strength of the asymmetry.

\section{Results and discussion}
\label{res}

\subsection{The energy dependence} 

Although the recent data \cite{DLHCb} are given in terms of the pseudo-rapidity $\eta$, in order to study the
energy dependence it is more convenient to use the Feynman $x_F$ variable, which may be written as:
$$
x_F= \frac{2 \, m_T \, cosh (y)}{\sqrt{s}} \simeq \frac{2 \, m_T \, cosh (\eta)}{\sqrt{s}}
     \simeq \frac{2 \, m_T \, e^{\eta}}{\sqrt{s}}
$$
for $\eta > 1$. Moreover, if the transverse momentum of the final $D$ meson is zero or
very small, then $m_T \simeq m_D$.

In order to study the energy dependence of the asymmetry, we shall focus on the two energies 
where we have experimental data: $\sqrt{s_1}= 33$ GeV and $\sqrt{s_2}= 7000$ GeV and construct the 
asymmetry ratio:
\begin{equation}
R_A \equiv  \frac{A(\sqrt{s_2})}{A(\sqrt{s_1})} = \frac{\Phi_D(\sqrt{s_2})/\Phi_T(\sqrt{s_2})}{\Phi_D(\sqrt{s_1})/\Phi_T(\sqrt{s_1})} \,\,\,.
\label{assrat}
\end{equation}
Using the definitions of the splitting function  (\ref{ftho}) in  (\ref{direct}) and then in (\ref{assym}), 
many energy independent factors cancel out and we find:
\begin{eqnarray}
R_A &=&  \frac{A(s_2)}{A(s_1)} = 
\left(\frac{\sigma^{\Sigma}(s_2)}{\sigma^{\Sigma}(s_1)} \right) / \left(\frac{\sigma_0^D(s_2)}{\sigma_0^D(s_1)} \right)  \nonumber \\
&=& \left(\frac{\sigma^{pp}(s_2)}{\sigma^{pp}(s_1)} \right) / \left(\frac{\sigma_{c \bar{c}}(s_2)}{\sigma_{c \bar{c}}(s_1)} \right)
\label{ratf1}
\end{eqnarray}
where in the last step we have used (\ref{estsig}) and  assumed that $\sigma^{\Sigma} = \sigma^{\Sigma^{++}_c \, p } 
= \mbox{const} \, . \,  \sigma^{p \, p }$. Moreover, we have neglected the energy dependence of $n_D$.
%\begin{equation}
%R_A = 
%\left(\frac{\sigma^{pp}(s_2)}{\sigma^{pp}(s_1)} \right) / \left(\frac{\sigma_{c \bar{c}}(s_2)}{\sigma_{c \bar{c}}(s_1)} \right)
%\label{ratf2}
%\end{equation}
The above ratios can be estimated with the recently obtained experimental data listed in Table \ref{tab1}.
\begin{table}[t]
%\begin{center}
\begin{tabular}{|c|c|c|}
\hline
Energy (GeV) & $\sigma_{p p}$ (mb) &  $\sigma_{c \bar{c}}$ (mb)\\  
\hline
\hline
33 & 40 & 0.04   \\
\hline
7000 & 97 & 8  \\
\hline
14000 & 110 & 11 \\
\hline
\end{tabular}
\caption{\small Total $p p$ cross section from \cite{fag} and total $c \bar{c}$ production cross section from \cite{ram} as a function of the 
energy. The first two lines refer to measurements and the last one show model calculations described in the corresponding references. }
\label{tab1}
%\end{center}
\end{table}
\begin{center}
\end{center}
With these numbers we find that increasing the energy from $\sqrt{s} = 33$ GeV to $7000$ GeV the asymmetry ratio is $R_A = 1/75 $, i.e., 
there is a strong decrease in the asymmetry. This 
happens because  meson emission (which ultimately causes the asymmetry) is  a non-perturbative process and has a slowly growing cross section. 
In contrast, the symmetric processes are driven by the perturbative partonic interactions, which have strongly growing   cross sections. 
We end this subsection making the prediction for the order of magnitude of the $D^+/D^-$ asymmetry  in the forthcoming $14$ TeV $p p$ collisions. 
Using the last line of Table \ref{tab1},  setting $\sqrt{s_2} = 14 $ TeV and susbtituting the numbers in (\ref{ratf1}) we find 
$R_A = 1/100 $ showing the decreasing trend of the asymmetry.  
\subsection{Predictions for the  asymmetries }
%Connection between $D^+/D^-$, $D^0/\bar{D}^0$ and  $\Lambda_c/\bar{\Lambda}_c$ asymmetries
Due to the lack of experimental data a direct comparison of $D^+/D^-$ asymmetries in
the same reaction, e.g. proton-proton, at different energies is difficult. However, we
can relate and compare similar reactions.
\begin{figure}[t]
\includegraphics[scale=0.30]{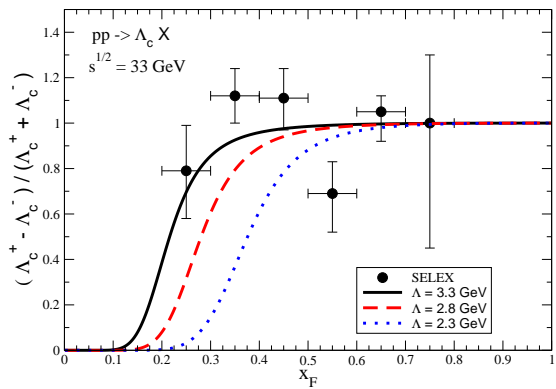}
\caption{Comparison of the MCM asymmetry for $\Lambda_c$ production  with experimental 
data \cite{selex}  for $\sqrt{s} = 33$ GeV.}
\label{Fig. 3}
\end{figure}
In what follows we shall relate the two sets of data on asymmetries in proton-proton collisions: the low energy one
taken by the SELEX collaboration \cite{selex} and the high energy one obtained by the LHCb collaboration \cite{DLHCb}.  
We will fit the data on  $\Lambda_c/\bar{\Lambda}_c$ asymmetry. This will  fix  the value of $N^{\Lambda}$ and 
define a range of possible values for the  cut off $\Lambda_{\bar{D}_0 p \Lambda_c}$. Using experimental information it is 
easy to go from the estimate of $N^{\Lambda}$ to the estimate of  $N^{D}$ and then it is straightforward to  calculate the 
associate asymmetry in  $\bar{D}_0 / D_0$. In proton-proton collisions the leading charm mesons are those with valence quarks and hence
$\bar{D}_0 (u \bar{c})$ has the same properties as $D^- (d \bar{c})$. From this approximate equivalence we infer the
$D^- / D^+$ asymmetry at the lower energy. The final step is to calculate this asymmetry at the LHC energy. This extrapolation 
can be done, since the model has a well defined energy dependence and fortunately the necessary ingredients
(the cross sections) are already available. 
\begin{figure}[t]
\includegraphics[scale=0.30]{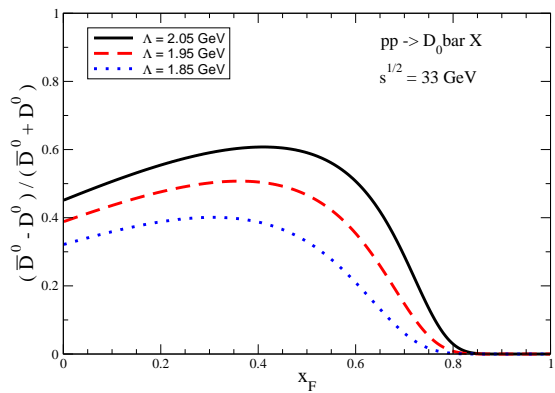}
%\vspace{0.2cm}
\caption{ MCM prediction of the $\bar{D}^0/D^0$  asymmetry, Eq.~(\protect\ref{finassytho}).}
\label{Fig. 4}
\end{figure}
\subsubsection{The coupling constants}
The coupling constant $g_{p D \Lambda_c}$   was estimated in several works with QCD sum rules \cite{gndl}. Here we have chosen a representative 
number.  In our analysis we will also need the coupling $g_{p D^- \Sigma^{++}_c}$, about which, 
to the best of our knowledge, nothing is known. As a first guess we will then assume that 
\begin{equation}
g_{p D^- \Sigma^{++}_c} = g_{p \bar{D}_0 \Lambda_c}  = 5.6 \,\,\,.
\label{coup}
\end{equation}
\begin{figure}[t]
\vspace{1.0cm}
\includegraphics[scale=0.30]{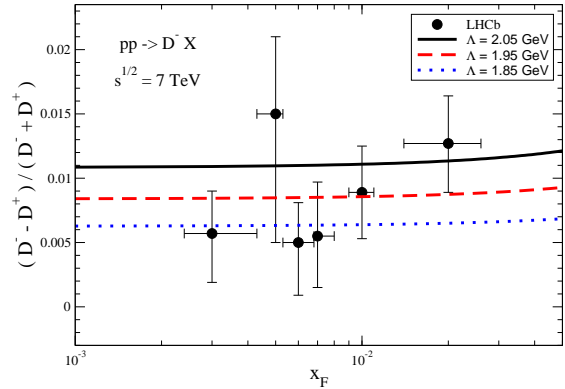}
%\vspace{0.2cm}
\caption{ Comparison of the MCM asymmetry, Eq.~(\protect\ref{finassytho}),
with experiental data \protect\cite{DLHCb} for $D^-/D^+$.}
\label{Fig.5}
\end{figure}
\subsubsection{The interaction cross sections of the cloud particles}
The  $D$ - proton and $\Lambda_c$ -  proton total cross sections in the $10 - 40$ GeV range are unknown. 
%We might know them better in the future from the CBM and PANDA experiments at FAIR.
We know that the presence of the heavy quark makes these states much more compact 
than the proton and hence with smaller geometrical cross section.  In the  well studied  case of the $J/\psi$ - proton cross section, a 
series of theoretical works gradually converged to  $\sigma^{J/\psi p}  \simeq 4$ mb \cite{noscross}. Moreover, it was found that, in contrast 
to the expectation of additive quark models, $\sigma^{J/\psi p} \simeq \sigma^{J/\psi \pi}$. Inspired by these previous works we shall 
assume that:
\begin{equation}
\sigma^{D p} \simeq \sigma^{\Lambda_c p} =  0.15 \, \sigma^{pp}   = 6 \, \mbox{mb} \,\,\,.
\label{chacross} 
\end{equation}
\begin{figure}[t]
\vspace{0.2cm}
\includegraphics[scale=0.30]{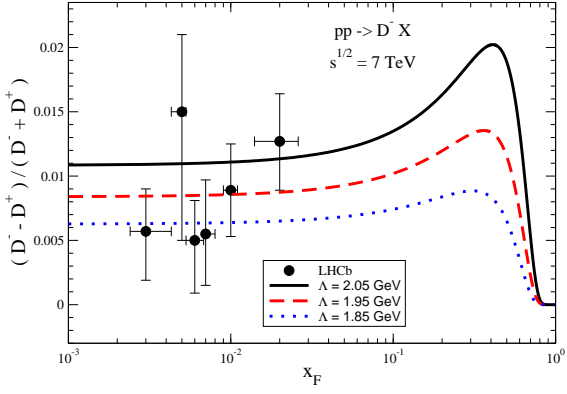}
%\vspace{0.2cm}
\caption{ Comparison of the MCM asymmetry, Eq.~(\protect\ref{finassytho}),
with experimental data \protect\cite{DLHCb} for $D^-/D^+$. Extension of Fig. \ref{Fig.5} to the large $x_F$ region.}
\label{Fig.6}
\end{figure}
\subsubsection{The production cross sections of the charm particles}
The production cross section of charmed hadrons is measured in some cases and calculated with pQCD in others. 
In order to have an estimate of these cross sections it is enough for our 
present purposes to use the pQCD results published in \cite{gavai}, from where we have deduced (\ref{estsig}) and (\ref{estlamzer}). 
All this information is summarized in Table \ref{tab2}.
\begin{table}[t]
\begin{tabular}{|c|c|c|c|c|}
\hline
$g_{p D \Lambda_c}$ & $ \sigma^{D p} $ & $ \sigma^{\Lambda}_0 $ & $\sigma^{\Lambda_c p} $ & $ \sigma^D_0 $  \\
\hline
%\hline
$5.6 $ & $ 0.15 $  $ \sigma^{pp} $  &  $ 0.15  $ $ \sigma_{c \bar{c}} $ & $ 0.15 $  $ \sigma^{pp} $  & $1.5$  $\sigma_{c \bar{c}}$ \\
\hline
\end{tabular}
\caption{\small Parameters used to calculate $N$ at $\sqrt{s} =33$ GeV.}
\label{tab2}
\end{table}
Starting from the definition (\ref{defn2}) and using the numbers given in  Table \ref{tab2} we have:
\begin{equation}
 N^{\Lambda} = \frac{g^2_{p D \Lambda_c}}{ 32 \, \pi}  \frac{ \sigma^{D p}} {\sigma_0^{\Lambda}} 
                 = \frac{(5.6)^2}{32 \, \pi} \frac{0.15 \, \sigma^{pp}}{0.15 \, \sigma_{c \bar{c}}} \simeq 320
\label{n2l}
\end{equation}
and also: 
\begin{equation}
 N^{D} = \frac{g^2_{p  D  \Lambda_c}} { 32 \, \pi}   \frac{ \sigma^{ \Lambda_c  p}} {\sigma_0^{D}}
                 = \frac{(5.6)^2}{32 \, \pi}    \frac{0.15 \, \sigma^{pp}} {1.5 \, \sigma_{c \bar{c}}}    \simeq 32
\label{n2D}
\end{equation}
\begin{figure}[t]
\vspace{0.2cm}
\includegraphics[scale=0.30]{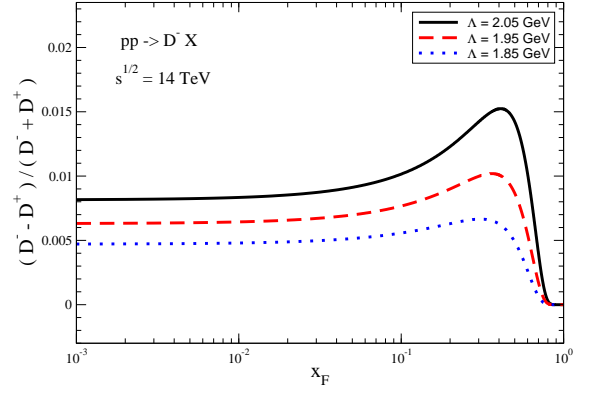}
%\vspace{0.2cm}
\caption{ Prediction  of the $D^-/D^+$  asymmetry for $\sqrt{s} = 14$ TeV.}
\label{Fig.7}
\end{figure}

In Fig. \ref{Fig. 3}, we show the $\Lambda_c/\bar{\Lambda_c}$ asymmetry in $p p $ collisions at $\sqrt{s} = 33$ GeV. The experimental
points are the results obtained by the SELEX collaboration. The  lines are calculated with Eq. (\ref{finassla} ) 
with the normalization fixed by (\ref{n2l}).   These data can not impose a stringent contraint
on the model, but they can establish a range of acceptable values for the cut-off parameter, which values are shown in the figure.
The outcoming numbers for $\Lambda$ are those expected in this kind of meson cloud  calculation. If they had been smaller than $1$ GeV 
or larger than $5$, this would have been an evidence against the model. 

In Fig. \ref{Fig. 4},  we show the  $\bar{D_0} / D_0$  asymmetry in $p p $ collisions at $\sqrt{s} = 33$ GeV. It was 
calculated with Eq. (\ref{finassytho}) with the normalization factor given by (\ref{n2D}). Since $N^D$ is fixed the only free parameter is the 
cut-off $\Lambda$, which, as it will be seen next, will be fixed so as  to yield a good fit of the $D^+/D^-$ asymmetry data from the LHCb 
collaboration. Also in this case, the values of  $\Lambda$ used to draw the curves are quite reasonable.  
The shapes of Figs. \ref{Fig. 3} and \ref{Fig. 4} are  correlated since they refer to 
the same vertex, where the proton splitts into a meson $\bar{D}_0$ and a baryon $\Lambda_c$. 
Due to its larger mass the baryon takes most of the momentum 
and the  resulting asymmetry peaks at very large values of $x_F$. Complementarily, the $\bar{D}_0$ distribution peaks at lower values 
of $x_F$, which, nevertheless, are still large.  The value of the cut-off $\Lambda$ does not have to be the same as that  used in Fig. \ref{Fig. 3} 
because, even though the coupling constant of a given vertex is always the same, the functional form of the  form factor (as a function of the 
off-shell particle squared momentum) changes when the off-shell particle changes. 
\begin{figure}[t]
%\vspace{0.2cm}
%\includegraphics[scale=0.35]{duplogamasohgg.eps}
\includegraphics[scale=0.30]{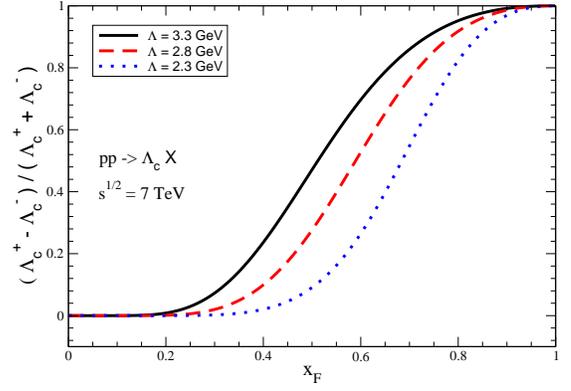}
%\vspace{0.2cm}
\caption{ Prediction  of the $\Lambda_c/\bar{\Lambda}_c$  asymmetry for $\sqrt{s} = 7$ TeV.}
\label{lambda_lhc}
\end{figure}

Neglecting differences coming from the different isospin, we assume that, apart from trivial changes in the masses,  
the vertex $ p \Sigma_c^{++} D^- $ has the same splitting function as the previously discussed $p \Lambda_c \bar{D}_0$ vertex and 
therefore the asymmetry of $D^-$ production is given by the same expression used for the $\bar{D}_0$.  Now we  try to reproduce the 
$\sqrt{s} = 7$ TeV LHCb data with  Eq. (\ref{finassytho}). The only part of this expression which depends 
on the energy is the factor $N^D$, which at higher energies will be corrected by the factor (\ref{ratf1}):
\begin{eqnarray}
N^{D}(\sqrt{s} = 7 \,  \mbox{TeV}) &=&  R_A \, . \, N^D   (\sqrt{s} = 33 \,  \mbox{GeV})  \nonumber \\
              &=&  \frac{1}{75}  \,   N^D   (\sqrt{s} = 33 \,  \mbox{GeV})
\label{reln}
\end{eqnarray}
In Fig. \ref{Fig.5} we show  Eq. (\ref{finassytho}) with $N^{D}(\sqrt{s} = 7 \, \mbox{TeV})$ and compare with the 
data, properly rewritten in terms of $x_F$ and with the definition (\ref{assym}), which introduces a minus sign with respect to \cite{DLHCb}.
In spite of the large error bars we can see that the MCM is able to reproduce the non-vanishing asymmetry. The data are surprisingly sensitive to 
cut-off choices, being able to discriminate small variations in $\Lambda$.  These data can not yet give a detailed information about the 
$x_F$ dependence of the asymmetry, but they show clearly that this asymmetry exists and also that it is much smaller than what we 
expect to find at lower energies for $p p $ and than what 
we have already found for pion and $\Sigma$ projectiles. 
After this close look into the data points, this figure deserves a zoom out to reach the 
$x_F$ region which was scanned in previous lower energies experiments.  This is shown in Fig. \ref{Fig.6}, from where we draw the most important 
conclusion of this work: the asymmetry definitely decreases at increasing energies, reaching at most 2 \% at $x_F \simeq 0.4$. 
Finally, in Fig. \ref{Fig.7} we show our prediction for the asymmetry to be measured at $\sqrt{s} = 14$ TeV.  For completeness, we present in Fig. \ref{lambda_lhc} our predictions for the $\Lambda_c$ asymmetry at $\sqrt{s} = 7$ TeV. 
\section{Summary}
\label{sum}
In this paper we have shown that the MCM provides a good understanding of the charm
production asymmetries in terms of a simple physical picture with few parameters.
It connects the behavior
of the asymmetries at large $x_F$ with the charm meson momentum
distribution within the cloud state. We can explain why we
observe asymmetries, why they are different for different beams and  why they decrease with increasing energies.  
\acknowledgments
This work has been partially supported by CNPq and  FAPESP.

\end{document}